Title:   Interlinked Dual-Time Feedback Loops can Enhance Robustness to Stochasticity and Persistence of Memory

Authors:   Paul Smolen, Douglas A. Baxter, and John H. Byrne

Laboratory of Origin:

Department of Neurobiology and Anatomy
W.M. Keck Center for the Neurobiology of Learning and Memory
The University of Texas Medical School at Houston
P.O. Box 20708
Houston, TX 77225

Correspondence Address:

Paul Smolen
Department of Neurobiology and Anatomy
W.M. Keck Center for the Neurobiology of Learning and Memory
The University of Texas-Houston Medical School
P.O. Box 20708
Houston, TX 77225
Voice: (713) 832-859-3826
FAX: (713) 500-0623
E-mail: Paul.D.Smolen@uth.tmc.edu

**ABSTRACT**

Multiple interlinked positive feedback loops shape the stimulus responses of various biochemical systems, such as the cell cycle or intracellular calcium release. Recent studies with simplified models have identified two advantages of coupling fast and slow feedback loops. Namely, this dual-time structure enables a fast response while enhancing resistances of responses and bistability to stimulus noise. We now find that in addition: 1) the dual-time structure confers resistance to internal noise due to molecule number fluctuations, and 2) model variants with altered coupling, which better represent some specific systems, share all the above advantages. We develop a similar bistable model with a fast autoactivation loop coupled to a slow loop, which minimally represents positive feedback that may be essential for long-term synaptic potentiation (LTP). The advantages of fast response and noise resistance carry over to this model. Empirically, LTP develops resistance to reversal over ~1 h. The model suggests this resistance may result from increased amounts of synaptic kinases involved in positive feedback.

# INTRODUCTION

Many biological systems have positive feedback as a core regulatory element that generates steep, or even switch-like, responses to graded stimuli [1, 2, 3, 4, 5, 6]. Well-studied cases include inositol 1,4,5,-trisphosphate (IP3) – induced release of $Ca^{2+}$ from the endoplasmic reticulum (ER) [7], the triggering of cell mitosis [8, 9], and maturation of *Xenopus* oocytes [10, 11]. Often, dual positive feedback loops reinforce each other. For example, rapid positive feedback in which $Ca^{2+}$ enhances its own release from the ER is reinforced by a slower rise in cytosolic $Ca^{2+}$ due to plasma membrane influx. This influx is mediated by store-operated $Ca^{2+}$ channels activated following depletion of ER $Ca^{2+}$ [12, 13]. Multiple positive feedback loops have been postulated to contribute to the formation and maintenance of long-term potentiation (LTP) of synaptic connections. Similarly, in invertebrates, multiple positive feedback loops have been posited to contribute to long-term facilitation of synapses (LTF) [14]. In the mollusc *Aplysia*, a positive feedback loop involving the transcription factor CREB1 plays an essential role in LTF [15].

Two recent studies have used simple, generic models to examine how organisms might gain an advantage by using reinforcing positive feedback loops. Brandman et al. [2] considered a two-loop monostable model in which species A and B cooperate additively to enhance production of an output species $C_{out}$. $C_{out}$ feeds back to increase the production of both A and B. For the case of one loop fast, with a small time constant for A to adjust to changes in $C_{out}$, and the second loop slow, with a slow time constant for B to adjust to $C_{out}$, two advantages were found. The fast loop enabled a rapid response, with $C_{out}$ rising quickly after stimulus. The slow loop increased the robustness of response amplitude and shape. Its slow time constant filtered out stimulus fluctuations, decreasing fluctuations in $C_{out}$. Such a fast-loop – slow-loop arrangement is termed a <u>dual-time</u> system. Following stimulus removal, a slow turn off of $C_{out}$ production was in this monostable model observed, governed by the slow time constant of the B variable.

Zhang et al. [16] studied a similar, dual-time model that exhibits bistability and hysteresis due to stronger positive feedback from $C_{out}$ to the synthesis of A and B. $C_{out}$ remains elevated after stimuli are terminated. Such a bistable model represents a switch in which a brief stimulus can cause a persistent state change. The fast positive feedback loop was again found to drive a rapid stimulus response, and the slow loop increased the stability of the basal and elevated states against stimulus fluctuations. In gene regulation, bistable switches have been hypothesized to convert brief stimuli into long-lasting state changes, such as cellular differentiation [4] or persistent gene activation [6, 17].

In the above models, A and B add together to increase production of $C_{out}$. Similar generic models can describe interlocking feedback loops with other topologies. For example, A and B may multiply to increase production of $C_{out}$. This multiplicative case seems to better describe some biochemical systems (see Discussion). The present study examines ways in which multiplicative production of $C_{out}$ affects responses to stimuli. In addition, the present study investigates the robustness of stimulus responses and steady states against internal system noise (stochastic fluctuations in the copy numbers of molecules). Internal noise can destabilize a bistable system, causing random jumps between states [18, 6]. Brandman et al. [2] and Zhang et al. [16] did not examine whether a dual-time architecture confers resistance to internal noise.

Additive and multiplicative production of $C_{out}$ share a common "convergent" topology (Fig. 1A, A and B converge to produce $C_{out}$). We also examined the dynamics of a distinct, but similar topology in which the fast variable A enhances its own formation and also that of a slow variable B. B further enhances formation of A. This "autoactivation" topology is motivated by positive feedback postulated to contribute to LTP and/or LTF, in which specific kinases (CAMKII, MAPK) enhance their own activity directly or indirectly. Some posited loops involved in LTP or LTF, dependent on kinase phosphorylation and activation, are likely to have more rapid time constants than other loops dependent on translation and on transcription (see Results and Discussion for details of loops). Therefore this autoactivation model is dual-time. The fast variable A corresponds to the level of active kinase and the slow variable B corresponds to the level of total kinase.

For the models of Brandman et al. [2] and Zhang et al. [16] with feedback modified so that A and B multiply to increase $C_{out}$, we find that the dual-time architecture again confers resistance to stimulus noise. Following a stimulus, either a fast or slow turn on of $C_{out}$ production could be obtained, depending on parameters. With the model of Zhang et al. [16], robustness to internal noise was enhanced by the dual-time architecture. As the time constant of one loop was increased, the average time required for molecule number fluctuations to destabilize a steady state increased rapidly. With the autoactivation model, the advantages of a) a rapid response to stimuli, and b) resistance of bistability to stimulus noise were again present. Robustness to internal noise was enhanced by the slow loop. Following a brief imposed kinase activation, the total kinase amount increased to a new plateau, at which there was only one stable, elevated, solution for kinase activity. Subsequent brief stimuli could not induce a state transition to low kinase activity. These dynamics suggest an explanation for development of resistance of LTP to depotentiation (see Discussion).

**METHODS**

For simulations with no explicit, external noise sources (Figs. 2, 5) the forward Euler method was used for integration of differential equations, with a time step of 5 msec. Simulations verified further time step reductions did not significantly improve accuracy. To further verify accuracy, the simulation of Fig. 5B was repeated using the second-order Runge-Kutta integration method [19]. No significant differences were observed. Prior to any stimulus, variable values were determined by equilibration for at least one simulated day, establishing steady-state levels of concentrations or molecule numbers. Longer equilibrations did not significantly alter these levels. The model was programmed in Java and simulated on Pentium 3 microcomputers. Programs are available at http://nba.uth.tmc.edu/homepage/jbyrne/assets/code/TwoLoopCode.zip.

Bifurcation analysis examined how steady-state levels of variables (A, B, $C_{out}$) depend on the strength of a constant applied stimulus that acts to increase the rate of production of A and B. The bifurcation software MATCONT was used (available at http://www.matcont.ugent.be).

Stimulus noise was simulated substantially as in Zhang et al. [16]. A white Gaussian noise term with mean zero was added to the deterministic stimulus (variable S). The standard deviation was 15-20 percent of stimulus amplitude, with specific values given in the text or figure legends. Fluctuations that took S to negative values were reset to S = 0. Ordinary differential equations were used with the noisy stimulus term. The Box-Mueller algorithm [19] generated a Gaussian term at each time step for which the noise was updated. The noise term had the form $S = S_0 + \sigma \sqrt{-2 \ln(U_1)} \cos(2 \pi U_2)$ where $U_1$ and $U_2$ are uniformly distributed random numbers. A point not discussed in Brandman et al. [2] or Zhang et al. [16] is that fluctuation amplitudes for model variables depend strongly on the chosen time step between updates of the noise term. To yield significant fluctuations in the fast variable A but not in the slow variable B, the noise time step must be small relative to the time constant of B but not relative to the time constant of A. For Fig. 1B, the noise time step was 1 s, satisfying this condition.

For stochastic simulations, fluctuations in the copy numbers of A, B, and $C_{out}$ were simulated with the Gillespie algorithm. This algorithm takes variable time steps, and during each time step, exactly one reaction occurs. Which type of reaction occurs is determined randomly, with the probability of each reaction type proportional to its deterministic rate expression. For further details see Gillespie [20, 21]. In Eqs. 1-9, each term on the right-hand side corresponds to a distinct deterministic reaction rate. These rates were used directly in the Gillespie algorithm. In stochastic simulations, the molecule numbers are scaled by using a volume factor $\Omega$. Increasing $\Omega$ corresponds to increasing volume while keeping average copy numbers per unit volume the same. Zero-order rate constants, such as basal rates of synthesis of a molecule, are multiplied by $\Omega$, as are Michaelis or Hill constants. First-order rate constants are not changed. Second-order rate constants are divided by $\Omega$. Fixed upper bounds for molecule numbers are multiplied by $\Omega$. $\Omega$ has units of $\mu M^{-1}$ to convert concentration to molecule number.

**RESULTS**

*Dual-time, multiplicative positive feedback exhibits stimulus noise resistance and variable response kinetics*

The models of Brandman et al. [2] and Zhang et al. [16] are schematized in Fig. 1A. The equations are as follows:

Brandman et al. [2] (henceforth denoted **Mod-B05**):

$$\tau_A \frac{dA}{dt} = \left[ S \frac{C_{out}^3}{C_{out}^3 + K^3} \right](1-A) - A + k_{min} \qquad (1)$$

$$\tau_B \frac{dB}{dt} = \left[ S \frac{C_{out}^3}{C_{out}^3 + K^3} \right](1-B) - B + k_{min} \qquad (2)$$

$$\frac{dC_{out}}{dt} = k_{on}(A+B)(1-C_{out}) - k_{off}C_{out} + k_{minout} \qquad (3)$$

Concentration units of $\mu M$ and time units of s are used. The following standard parameter values are used unless noted in the figure legends:

$\tau_A = 2.0$ s, $\tau_B = 125.0$ s, $K = 0.35$ $\mu M$, $k_{min} = 0.01$ $\mu M$,
$k_{on} = 2.0$ $\mu M^{-1} s^{-1}$, $k_{off} = 0.3$ s$^{-1}$, $k_{minout} = 0.001 \mu M$ s$^{-1}$

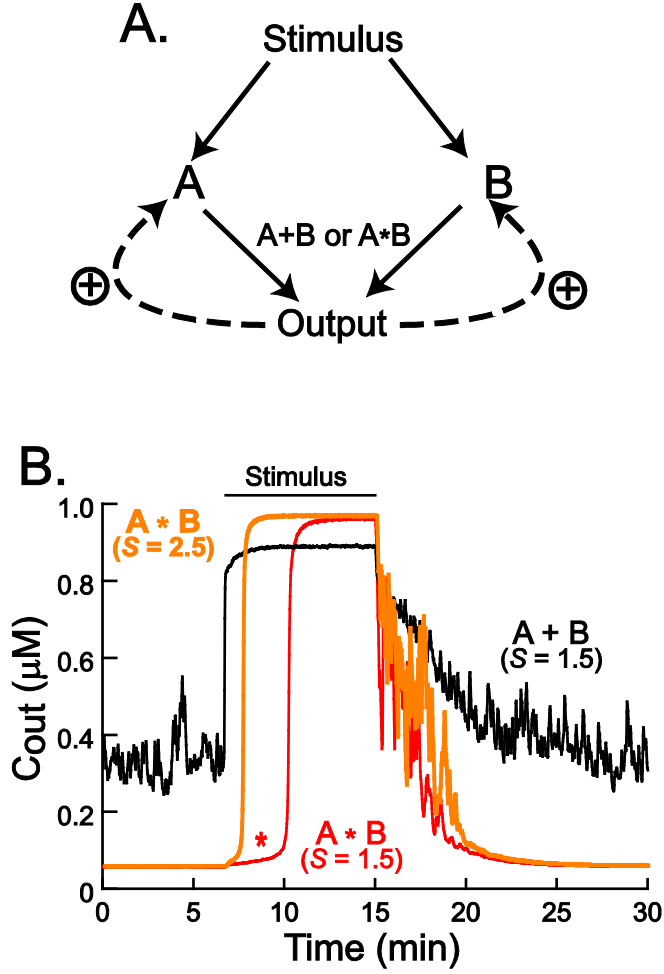

**Figure 1.** (Color Online) Dynamics of coupled fast and slow positive feedback loops. (A) Schematic of the coupling in the models of Brandman et al. [2] and Zhang et al. [16]. The rate of synthesis of OUT is proportional to the sum of A and B. OUT feeds back to enhance the synthesis of A with a fast time constant $\tau_A$, and enhances the synthesis of B with a slow time constant $\tau_B$. This schematic also describes multiplicative coupling (Eq. 7). (B) Response to a stimulus pulse for the model of Eqs. 1-3 (OUT synthesis depends on A+B) or the model of Eqs. 1, 2, and 7 (OUT synthesis depends on A*B). Stimulus amplitude (parameter S) is zero until $t$ = 6.7 min, at which time S is set to either 1.5 or 2.5. At $t$ = 15 min S is reset to zero. Gaussian stimulus noise is always present with a standard deviation of 0.15 and an update time step of 1 s. Model parameters have the standard values given after Eqs. 1-3 and Eq. 7.

Zhang et al. [16] (henceforth denoted **Mod-Z07**):

$$\tau_A \frac{dA}{dt} = \left[ k_1 S + k_2 \frac{C_{out}^4}{C_{out}^4 + K^4} \right] (1-A) - A + k_{min} \tag{4}$$

$$\tau_B \frac{dB}{dt} = \left[ k_1 S + k_2 \frac{C_{out}^4}{C_{out}^4 + K^4} \right] (1-B) - B + k_{min} \tag{5}$$

$$\frac{dC_{out}}{dt} = k_{on} (A+B)(1-C_{out}) - k_{off} C_{out} + k_{minout} \tag{6}$$

Standard parameter values are:
$\tau_A = 2.0$ s, $\tau_B = 200.0$ s, $k_1 = 0.1$, $k_2 = 0.3$, $K = 0.5$ $\mu M$,
$k_{min} = 0.01$ $\mu M$, $k_{on} = 1.0$ $\mu M^{-1} s^{-1}$, $k_{off} = 0.3$ $s^{-1}$, $k_{minout} = 0.003$ $\mu M$ $s^{-1}$

Mod-Z07, but not Mod-B05, is bistable. Bistability in Mod-Z07 is due to greater nonlinearity in the feedback from $C_{out}$ to A and B [16]. Hill coefficients describing activation of A and B synthesis by $C_{out}$ are 4 in Mod-Z07 *vs*. 3 in Mod-B05.

For multiplicative positive feedback, Eqs. 3 and 6 are replaced by

$$\frac{dC_{out}}{dt} = k_{on}(A\,B)(1-C_{out}) - k_{off}C_{out} + k_{minout} \qquad (7)$$

The multiplicative model variants are denoted **Mod-B05-Mult** and **Mod-Z07-Mult**. Figure 1A still schematizes these model variants. Differences in standard parameter values between these variants and the original models (Eqs. 1-6) are as follows:
Mod-B05-Mult, $k_{on} = 20.0\ \mu M^{-1}s^{-1}$, $k_{off} = 0.3\ s^{-1}$, $k_{minout} = 0.015\ \mu M\ s^{-1}$.
Mod-Z07-Mult, $k_{on} = 12.0\ \mu M^{-1}s^{-1}$, $k_{off} = 0.3\ s^{-1}$, $k_{min} = 0.02\ \mu M$.

Figure 1B illustrates stimulus responses for Mod-B05 and Mod-B05-Mult. S is 0 except between $t$ = 6.7 min and 15 min. These time courses of $C_{out}$ were computed with noisy stimuli, using a Gaussian noise term (see Methods). With dual-time feedback, the upper plateau of $C_{out}$ is resistant to noise (only small fluctuations in OUT occur). When $C_{out}$ has intermediate values not close to 0 or 1, larger fluctuations in $C_{out}$ are seen (during the turn off of $C_{out}$, and the lower state of Mod-B05). For Mod-B05-Mult, its lower state is close to 0, and only small fluctuations in $C_{out}$ are observed. Thus, both models exhibit resistance to stimulus noise when $C_{out}$ is close to its bounding values (0 or 1).

In Fig. 1B, Mod-B05 exhibits a fast turn on to the stimulus. The feedback loop in which $C_{out}$ activates A production is fast (time constant $\tau_A$ = 1 s) whereas the loop in which $C_{out}$ activates B production is slow ($\tau_B$ = 100 s). Rapid induction of A by stimulus drives the fast turn on of $C_{out}$. After stimulus removal, A falls rapidly, but B remains high for longer, maintaining high $C_{out}$. $C_{out}$ decays slowly as B returns to its basal value. Brandman et al. [2] suggested Mod-B05-Mult should exhibit opposite dynamics from Mod-B05. With Mod-B05-Mult, a slow turn on to a stimulus pulse should be followed by a fast turn off after the stimulus. In Fig. 1B, for Mod-B05-Mult and S = 1.5, a biphasic turn on of $C_{out}$ is seen. An initial slow increase from $t$ = 7 min to 10 min (denoted by * in Fig. 1B), is followed by a rapid increase. For a greater stimulus (S = 2.5), a faster increase of $C_{out}$ is seen. Thus, for Mod-B05-Mult, the kinetics of $C_{out}$ induction vary substantially with the stimulus. Because induction is fast for a strong stimulus, these kinetics are not in general opposite to the fast turn on of $C_{out}$ seen with Mod-B05. However, the turn off of $C_{out}$ after stimulus removal is consistently rapid.

Figure 2A illustrates bifurcation diagrams of bistability for Mod-Z07 and Mod-Z07-Mult. For each diagram (A+B for Mod-Z07, A*B for Mod-Z07-Mult) there is a range of stimulus strength supporting two stable solutions for the concentrations of $C_{out}$, A, and B. For each diagram, the bistable range of S is between the knees, or limit points (LP). The upper and lower steady states of $C_{out}$ are stable to small perturbations and are separated by an unstable middle steady state. Mod-Z07-Mult tends to support a broader range of bistability. To examine the resistance of the upper and lower steady states to stimulus noise, Mod-Z07 and Mod-Z07-Mult were subjected to Gaussian stimulus noise. The mean value of S was 0.14 (within the bistability region of Fig. 2A), the standard deviation was 0.15 (same as in Fig. 1B), and the noise update time step was 1 s. Both the upper and lower states of Fig. 2A remained stable, for Mod-Z07 and Mod-Z07-Mult.

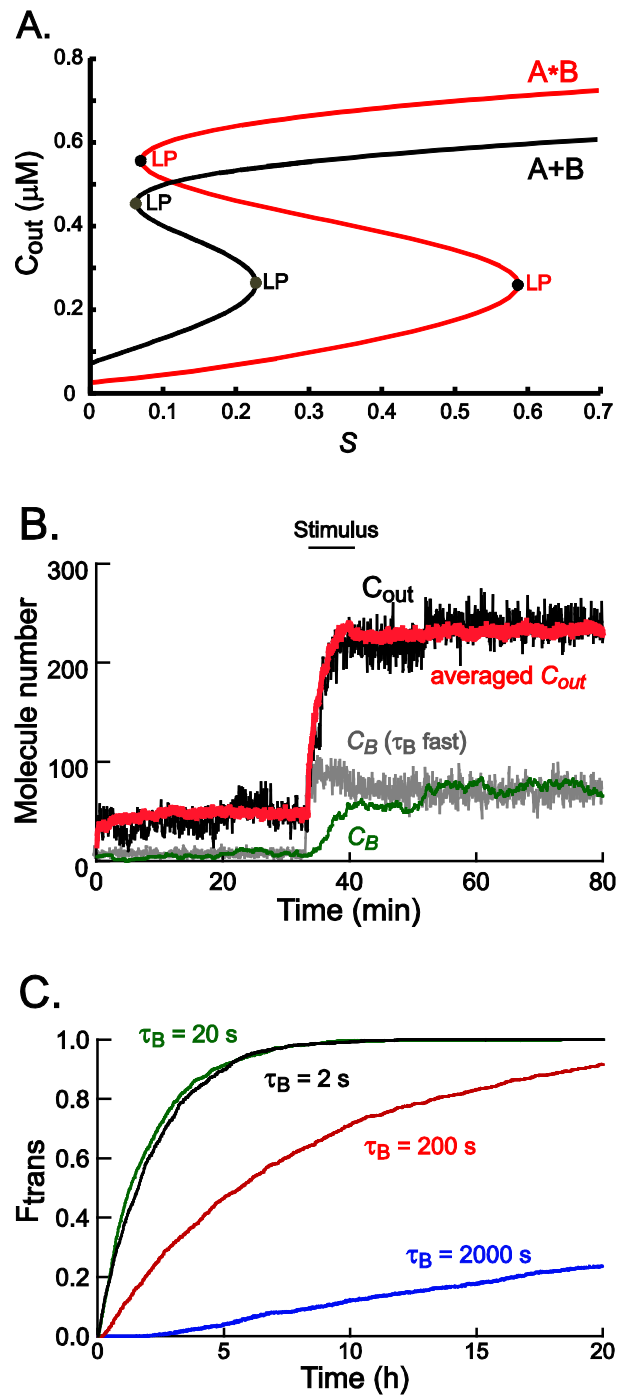

**Figure 2.** (Color Online) Bifurcation in the model of Zhang et al. [16]. (A) Bifurcation diagrams for the bistable variants of the model of (Fig. 1A). Steady states of $C_{out}$ are traced as a function of a constant value of S. Eqs. 4-6 with their standard values were used to compute the "A+B" curve, and Eqs. 4, 5, and 7 were used for the "A*B" curve. "LP" denotes a limit point at which a steady state vanishes. Parameters have the standard values. (B) Bistability in the model of Zhang et al. [16] is preserved with internal stochastic noise. Eqs. 4-6 were used. The volume factor $\Omega = 400$. Prior to choosing $\Omega$, all parameters were at standard values except $k_{on} = 1.2 \ \mu M^{-1} s^{-1}$, $k_{min} = 0.005 \ \mu M$, $k_{minout} = 0.001 \ \mu M \ s^{-1}$. Zero-order rate constants ($k_{min}$, $k_{minout}$) and the Hill constant K are multiplied by $\Omega$. The upper bounds for A, B, and $C_{out}$, which are 1.0 in Eqs. 4-6, are multiplied by $\Omega$. The second-order rate constant $k_{on}$ is divided by $\Omega$. Bistability for $\tau_A$ fast (2 s) and $\tau_B$ slow (200 s). The model is initialized in the lower state with S = 0.1. At $t$ = 33 min, S is increased to 0.5 for 7 min. The model transits to the upper steady state. The "average OUT" time course is over 20 simulations. For B, two time courses are shown. Time courses of B are shown with $\tau_B$ slow (200 s) and with $\tau_B$ fast (2 s). (C) Simulations of the time course of fluctuation-induced escape from the lower state to the upper state. Each time course represents the evolution of the fraction $F_{trans}$ of simulations that have transited at least once to the upper state, for an ensemble of 1,000 simulations. Model parameters as in (B) except $\tau_B$ varies.

*Dual-time feedback loops confer resistance of stimulus response to internal stochastic noise*

Stochastic fluctuations in molecule copy numbers are an ubiquitous source of internal noise in biological systems. Does a dual-time architecture confer resistance to this noise? We examined whether bistability in Mod-Z07 is robust to internal noise for time-average molecule numbers of ~100-300. The Gillespie algorithm was used (Methods). Average molecule numbers vary proportionately with the volume factor $\Omega$. Figure 2B illustrates that for $\Omega = 400$, lower and upper steady states are stable to internal noise. The "averaged $C_{out}$" time course is the average of the stimulus response over 20 simulations. In each simulation, S is at its basal value, 0.1, until $t$ = 33 min. At $t$ = 33 min, S was increased to 0.5, returning to 0.1 at $t$ = 40 min. S = 0.5 is to the right of the bistability range for Eqs. 4-6. Therefore, the model transits to the upper steady state.

In all 20 simulations, the system started in a stable, fluctuating lower state and ended in a stable higher state. However, stability of states was obtained for both a fast A loop and a slow B loop (time course of $C_{out}$ and slower time course of B, $\tau_A = 2.0$ s, $\tau_B = 200.0$ s) and for both loops fast (superimposed time course of B with faster fluctuations, $\tau_A = \tau_B = 2.0$ s). Thus these simulations did not demonstrate that a slow B loop conferred additional stability.

The stability of steady states as a function of $\tau_B$ was explored further, using ensembles of simulations similar to those in Zhang et al. [16]. Mod-Z07 was initialized in the low state. A constant, relatively low stimulus was applied (S = 0.14). Bifurcation analysis demonstrated that a stable upper state exists for this value of S. For an ensemble of 1,000 simulations, the time evolution of the fraction of systems that underwent a spontaneous, fluctuation-induced transition to the upper state was followed. On a time scale of hours, this fraction $F_{trans}$ exponentially rose toward 1. Figure 2C plots the time courses of $F_{trans}$ as a function of the time constant $\tau_B$. For $\tau_B$ fast (2 s) or intermediate (20 s), $F_{trans}$ rises relatively quickly. But for larger values of $\tau_B$, the increase in $F_{trans}$ is much slower, demonstrating a substantial increase in the stability of the lower state. In a complementary set of simulations, with S = 0.14 and with initialization of molecule numbers in the upper state, the stability of the upper state also increased with $\tau_B$ (not shown). Thus, these ensemble simulations succeeded in demonstrating greater stability of both steady states when the B loop was slow.

*Parallel unlinked feedback loops do not give both fast response and noise resistance*

We considered the extent to which coupling of loops is important for the response properties and noise resistance. A fast feedback loop in which A activates its own production was placed parallel to a slow feedback loop in which B activates its own production. As above, the rate of production of $C_{out}$ is driven by either the sum or the product of A and B. However, $C_{out}$ does not influence the production of either A or B. The model is schematized in Fig. 3A. For the case where production of $C_{out}$ is driven by a weighted sum of A and B, the equations are:

$$\tau_A \frac{dA}{dt} = S\,A\,(1-A) - A + k_{min} \qquad (8)$$

$$\tau_B \frac{dB}{dt} = S\,B\,(1-B) - B + k_{min} \qquad (9)$$

$$\frac{dC_{out}}{dt} = k_{on}(\lambda_1 A + \lambda_2 B)(1 - C_{out}) - k_{off} C_{out} + k_{minout} \qquad (10)$$

Standard parameter values are $\tau_A = 2.0$ s, $\tau_B = 100.0$ s, $k_{minout} = 0.001 \mu M\,s^{-1}$, $k_{off} = 0.3\,s^{-1}$, $k_{min} = 0.01\,\mu M$, $k_{on} = 0.3\,\mu M^{-1}s^{-1}$. Values for S, $\lambda_1$, and $\lambda_2$ are provided below and and in the legend to Fig. 3.

Figure 3B illustrates the dynamics of A and B in response to an applied, square-wave increase of S from 0 to 1.5, with Gaussian stimulus noise included as in Fig. 1B. The fast variable A increases rapidly to a fluctuating plateau. Because A is fast, the noise in S drives large fluctuations in A. When S returns to 0, A returns very rapidly to basal values. Variable B increases much more gradually to a plateau, and because its time constant is much longer, the noise in S drives only small fluctuations in B.

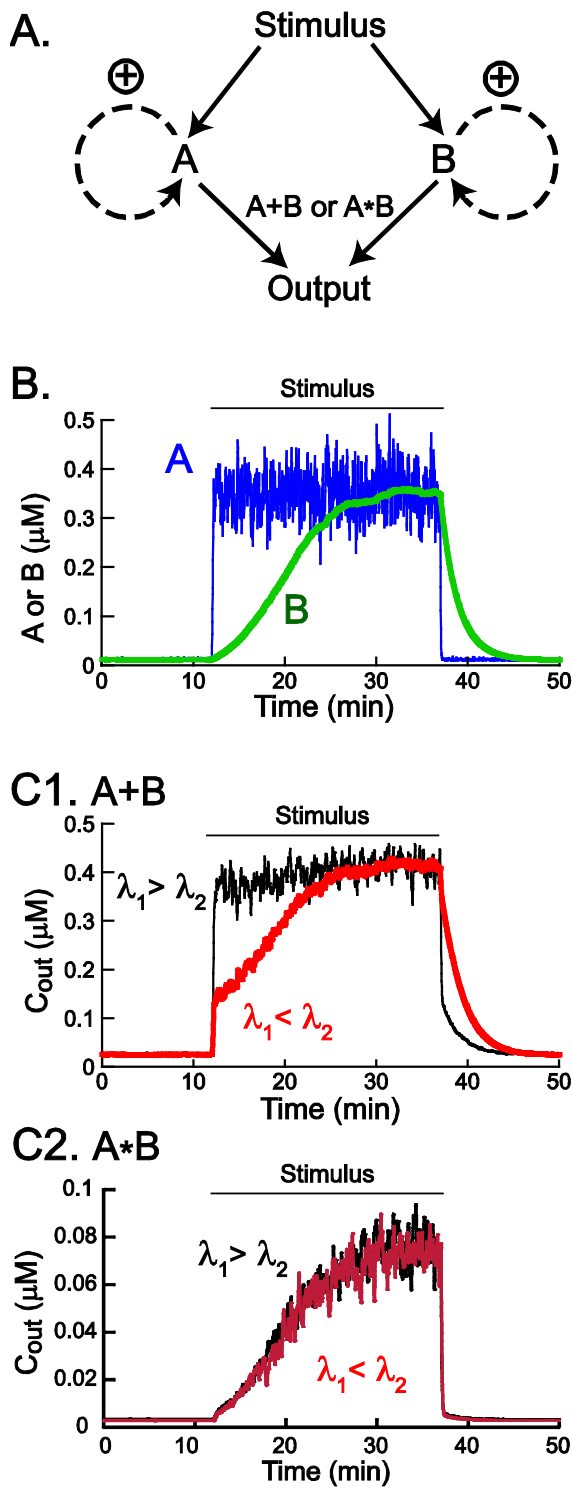

**Figure 3.** (Color Online) Dynamics of parallel uncoupled feedback loops. (A) Model schematic. The rate of synthesis of $C_{out}$ is proportional to either the sum or product of A and B. A feeds back to enhance its own synthesis with a fast time constant $\tau_A$, and B feeds back to enhance its own synthesis with a slow time constant $\tau_B$. (B) Simulated response of A and B to a stimulus. At $t = 12$ min, S is increased from 0 to 1.5, and remains elevated until $t = 38$ min, at which time S returns to 0. Gaussian stimulus noise is present with a standard deviation of 0.3 and an update time step of 1 s. A increases to a plateau with large fluctuations, whereas B increases slowly with small fluctuations. (C1) Simulated response of $C_{out}$ to the stimulus of (B) with A+B. Rapid time course, $C_{out}$ is driven mostly by changes in A ($\lambda_1 = 1.6$, $\lambda_2 = 0.4$). Slow time course, $C_{out}$ is driven mostly by changes in B ($\lambda_1 = 0.4$, $\lambda_2 = 1.6$). (C2) Simulated response of $C_{out}$ to the stimulus of (B) with A*B. The system is no longer sensitive to differential changes to the coupling strength of the two loops.

The strength of coupling between $C_{out}$ and A is given by the parameter $\lambda_1$ (Eq. 10), and the coupling between $C_{out}$ and B is given by $\lambda_2$. If $\lambda_1 > \lambda_2$, changes in $C_{out}$ will be predominantly driven by changes in A, whereas if $\lambda_2 > \lambda_1$, $C_{out}$ will mostly be driven by B. We considered two cases: 1) $\lambda_1 = 1.6$, $\lambda_2 = 0.4$, and 2) $\lambda_1 = 0.4$, $\lambda_2 = 1.6$. For $\lambda_1 > \lambda_2$, the response of $C_{out}$ to a stimulus is similar to that of A. A rapid increase, substantial fluctuations around a plateau, and a rapid decrease are observed (Fig. 3C1). For $\lambda_1 < \lambda_2$, the response of $C_{out}$ is similar to that of B, showing only small fluctuations (slow $C_{out}$ time course, Fig. 3C1). In neither case, are the dynamics similar to those of Fig. 1B or of Brandman et al. [2]. The combination of a rapid increase in $C_{out}$ and a slow decrease is not observed. When $C_{out}$ does increase rapidly, fluctuations in $C_{out}$ are substantial (slow time course, Fig. 3C1), so the combination of a rapid increase and resistance to noise is also not observed. We also considered the case with production of $C_{out}$ proportional to the product of A and B. For this case, noise resistance is also not obtained, because the slow B loop cannot damp noise in $C_{out}$. The use of the product A*B prevents the coupling of A to $C_{out}$ from being made small. Fluctuations in $C_{out}$ are driven by fluctuations in A even for constant B (Fig. 3C2).

The above simulations indicate that to obtain in concert the dynamic elements of a rapid increase in $C_{out}$, resistance to noise, and a slow decrease in $C_{out}$, the feedback loops cannot be uncoupled as in Fig. 3A and Eqs. 8-10. Instead, the loops must be coupled with $C_{out}$ feeding back to increase the production of A and B.

*A simple, bistable dual-time model represents aspects of the induction and consolidation of LTP*

LTP induction and consolidation has been proposed to involve positive feedback loops in which kinases such as CaM kinase II (CAMKII) or mitogen-activated protein kinase (MAPK) directly or indirectly enhance their own phosphorylation and activity. Persistent MAPK phosphorylation and activity might be maintained by reciprocal activation of MAPK and upstream Raf kinase [22, 23] or protein kinase C and MAPK [24, 25]. Self-sustaining phosphorylation and activation of CAMKII may occur [26, 27]. Inhibition of MAPK blocks LTP [28]. A feedback loop in which a kinase directly or indirectly enhances its own phosphorylation and activation can be generically represented with a variable A that activates its own production. A represents the amount of active kinase. The time scale of this loop is posited to be fast relative to transcription or translation. The total amount of kinase could be represented by a variable B, in which case the amount of active kinase A will be bounded by B. A differential equation for A can be written, similar to those in Zhang et al. [16], and representing autoactivation of A and the upper bound B:

$$\tau_A \frac{dA}{dt} = \left[ k_1 S + k_2 \frac{A^4}{A^4 + K^4} \right] (B - A) - k_{degA} A + k_{minA} \quad (11)$$

In a coupled, dual-time topology, a second slow positive feedback loop is posited in which active kinase A acts to increase total kinase B. The increase in total kinase would, in turn, tend to further increase the amount of active kinase *via* mass action. To argue that this loop is plausible, it is necessary to consider how an increase in active kinase could increase total kinase. MAPK can upregulate translation in neuronal processes [29, 30] as can CAMKII [31]. Indeed, CAMKII regulates the activity of CPEB, which in turn upregulates the synthesis of CAMKII during synaptic plasticity [32, 33, 34]. Thus, activation of CAMKII or MAPK could increase translation of proteins important for synaptic strengthening. MAPK may also phosphorylate transcription factors, increasing transcription of proteins important for synaptic strengthening (see Discussion). Levels of synaptic MAPK or CAMKII might therefore be increased by translation, transcription, or recruitment of preexisting kinase. Finally, increased levels and thus activity of MAPK or CAMKII would tend to further enhance local translation of synaptic proteins, thereby closing the positive feedback loop.

In the simplified model, the rate of increase of B is proportional to A. *In vivo* and in models of learning in neural networks, synaptic weights have upper bounds. In the model, an upper bound $B_{MAX}$ must be imposed on B to prevent A and B from increasing without limit. A differential equation for B that represents increase proportional to A and saturation at $B_{MAX}$ is as follows:

$$\tau_B \frac{dB}{dt} = k_3 A (B_{MAX} - B) - B + k_{minB} \quad (12)$$

LTP induction corresponds to a state transition for the level of A. Consolidation of LTP corresponds to a slow increase in B. The model is schematized in Fig. 4A. Standard parameter values are:

$\tau_A = 2.0$ s, $\tau_B = 3{,}600$ s, $k_1 = 0.1$, $k_2 = 1.0$, $K = 0.34$ $\mu$M, $k_{degA} = 1.0$

$k_{minA} = 0.08$ $\mu$M, $k_3 = 2.0$ $\mu$M$^{-1}$, $B_{MAX} = 4.0$ $\mu$M, $k_{minB} = 0.8$ $\mu$M

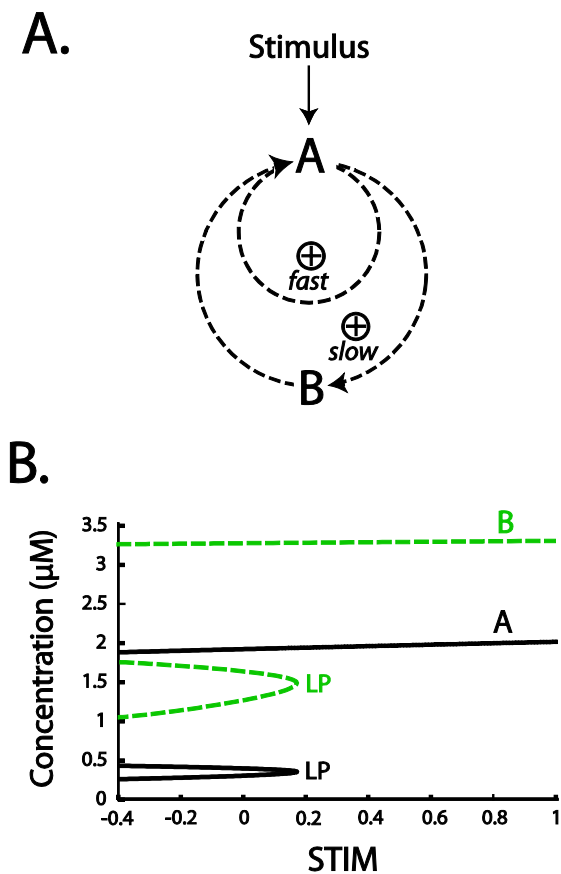

**Figure 4.** (Color Online) A bistable simplified model representing putative feedback loops in LTP and LTF induction and consolidation. (A) Model schematic. In a relatively fast feedback loop (time constant $\tau_A$), a kinase catalyzes (directly or indirectly) its own phosphorylation and activation. A denotes the amount of active kinase. An external stimulus, such as an influx of Ca$^{2+}$ into a neuron, would provide the initial kinase activation. In a second slow feedback loop (time constant $\tau_B$), A enhances the production of B which in turn promotes the formation of A. B could represent the total amount of kinase (inactive + active). (B) Bifurcation diagram illustrating the steady states of A and B as a function of a constant stimulus strength. Standard parameter values following Eqs. 11 and 12 are used.

Bifurcation diagrams illustrate a range of bistability for this model. Figure 4B shows that for constant stimulus amplitude ranging from negative values to ~ 0.2, stable lower and upper states of A and B coexist and are separated by a middle unstable steady state. In the model, S represents activation of signaling pathways such as the MAPK cascade, with S = 0 representing no activation. Thus the negative values of S are not considered physiological. As in previous models, the slow feedback loop (long $\tau_B$) confers resistance to stimulus noise. In an ensemble of 100 simulations with a noisy stimulus, the model was initialized in the lower steady state. A noisy stimulus with a constant mean of 0.15 was applied, with model parameters as in Fig. 4B. Gaussian noise with a standard deviation of 30 percent of the mean was applied. The time step for noise update was 1 s. This noise destabilized the lower state, but the resistance to destabilization increased rapidly with increases in $\tau_B$. For $\tau_B$ fast (1 s), the time for half of the ensemble simulations to transit out of the lower state ($t_{0.5}$) was only 0.1 h. For $\tau_B = 10$ s, $t_{0.5}$ was 1.0 h. For $\tau_B = 100$ s and for $\tau_B = 1{,}000$ s respectively, only 11 out of 100 and 3 out of 100 simulations were destabilized by noise during 6 h.

*Separation of fast and slow variables illustrates the way in which the upper state of the autoactivation model acquires resistance to reversal by brief stimuli*

Figure 5A illustrates that following a state transition of the model of Fig. 4 from the lower to the upper state of A, the upper state becomes more stable with time. To induce the transition, the value of S was briefly increased from its baseline of 0 to 200, for 1 s. The abrupt increase in A is followed by a gradual increase in both A and B over the next 2 h, due to the slow positive feedback loop with $\tau_B = 1$ h. At $t = 5$ h, $k_{degA}$, the degradation rate constant for A, was briefly increased to try to force A back to a lower state. Although A was driven to nearly zero, it immediately recovered to the upper state after $k_{degA}$ returned to its basal value. In contrast, if the same brief increase in $k_{degA}$ is applied earlier, 20 min after the upward transition in A, A is forced back to a lower state. At that time B is less, so that the lower state of A is stable. After return of A to the lower state, B declines to its lower state. Further simulations quantified the time required for the upper state of A to develop resistance to reversal. No reversal occurred if the brief decrease in $k_{degA}$ was applied more than 46 min after the upward transition in A.

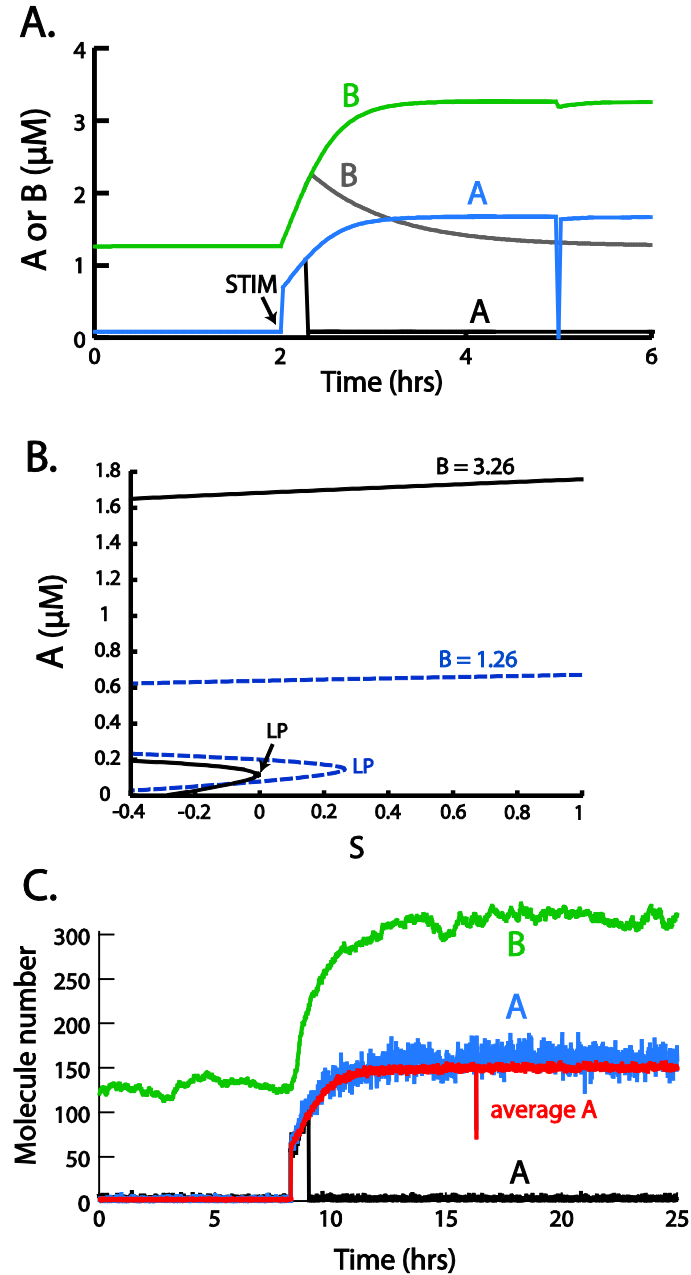

**Figure 5.** (Color Online) Resistance of elevated kinase activity to reversal by brief stimuli. (A) Development of resistance over time. At $t = 2$ h, a 1 s elevation of S to 200 induces a transition of A to the upper state. B increases over the next few hrs. At $t = 5$ h, a 100 s elevation of $k_{degA}$ from 1 to 11 fails to induce a state transition (upper B and A traces). A drops briefly but returns immediately to the upper state. If the same brief elevation of $k_{degA}$ occurs 20 min after the transition to the upper state, then A does transit back to the lower state, after which B slowly declines (lower B and A traces). Model parameters are at standard values. (B) Bifurcation diagrams for A as a function of S. The variable B is treated as a fixed parameter, thus only Eq. 11 is used to compute the diagrams. Standard parameter values following Eqs. 11 and 12 are used. Two bifurcation curves each show upper, middle, and lower steady states of A, for B = 3.26 and B = 1.26 respectively. (C) Stochastic simulation of resistance to depotentiation. $\Omega$ was chosen as 100. At $t = 8$ h a 1 s increase of S to 200 drives an upwards state transition. The "average A" trace is over 20 simulations. Superimposed is a time course of A for a single simulation. For all simulations, both the lower and upper state are stable. Parameter values differ somewhat from Fig. 5A because the lower steady state of Fig. 5A is not stable against internal noise for $\Omega \sim 100$. The changed parameter values are:

$\tau_B = 3$ hr, $K = 0.3$ $\mu$M, $k_{minA} = 0.018$ $\mu$M, $B_{MAX} = 3.6$ $\mu$M, $k_{minB} = 1.2$ $\mu$M

At $t = 17$ h, a 60 s increase of $k_{degA}$ from 1 to 3 fails to induce a state transition. However, the same brief increase in $k_{degA}$ applied shortly after the up transition returns A to the lower state.

Empirically, following induction of LTP *in vitro*, subsequent electrical stimuli can reverse LTP only when applied within ~ 1 h [35, 36]. The time course of ~ 1 h could correspond in the model to the time required for increases in the total amounts of synaptic proteins. Bifurcation analysis was used to examine whether dynamics of the slow variable B explain the development of resistance to state reversal. B was treated as a parameter in determining the steady states of the fast variable A. This type of fast-slow variable separation is often used to examine how dynamics of fast variables are altered gradually by changes in slow variables [37, 38, 39]. Figure 5B illustrates that with B fixed at 1.3, both lower and upper stable states of A exist for positive values of S (S < 0.26). As B is increased, the rate of synthesis of A increases (Eq. 11). With B increased to 3.3, the lower stable state of A only exists at negative values of S. In Fig. 5B, at $t$ = 5 h, B = 3.3. No stable lower state of A exists for physiological (nonnegative) S. Therefore, the upper state of A cannot be destabilized by brief stimuli. The transition of A to the upper state has become irreversible due to the subsequent increase in B.

For LTP induction and maintenance, many of the important biochemical processes occur within dendritic spines, which have volumes on the order of 0.1 femtoliters (fl) [40]. For these small volumes, molecule copy numbers are limited and internal noise is likely to be important. The dynamics of Fig. 5A were preserved when internal noise was simulated using the Gillespie algorithm. We chose $\Omega$ such that the average copy number of B is ~100-300. These copy numbers correspond to concentrations of 2-5 μM in 0.1 fl. Figure 5C illustrates the stability of steady states and the development of resistance of the upper state to reversal. The "average A" time course is over 20 simulations. Stability of upper and lower states was preserved for all 20 simulations. Resistance to state reversal developed by $t$ = 17 h. The downward spike shows that the upper state was resistant to a brief increase in $k_{degA}$ for all 20 simulations. In contrast, another time course shows that if the increase in $k_{degA}$ was applied soon after the upward transition, before B increased much, the upper state was not resistant. A was driven back to the lower state.

*Dual-time feedback loops confer resistance of bistability to internal noise, but increasing system volume has a greater effect*

With the autoactivation model, is the robustness of bistability increased when the feedback loop with B is slow (large $\tau_B$)? Using the method of Fig. 2C, we examined the time required for stochastic fluctuations to induce a transition from the lower state to the upper state. S was set to 0.1, $\Omega$ to 100, and for ensembles of 1,000 simulations, the model was initialized in the lower steady state of A and B. In Fig. 6A, each time course represents the increase over time of the fraction of simulations $F_{trans}$ that have undergone a transition to the upper state. The stability of the lower state increases with $\tau_B$. However, the increase is not as large as in Fig. 2C. When $\tau_B$ is increased above 100 s, little additional stability is seen. For $\tau_B$ = 1,000 s, the rate of increase of $F_{trans}$ is only slightly smaller than for $\tau_B$ = 100 s.

In two control ensemble simulations we verified that the upper and lower states were very stable in the absence of stimulus (S = 0). Over 24 h, with $\tau_B$ = 3,600 s, only 1 percent of 1,000 simulations transited out of the lower state. When both A and B were initialized at elevated values (A = 150, B = 300), none of 1,000 simulations transited to the lower state over 24 h.

Figure 6B illustrates that the volume factor $\Omega$ is a much stronger determinant of the robustness of bistability than is $\tau_B$. The time courses illustrate that the rate of increase of $F_{trans}$ decreases rapidly as $\Omega$ increases.

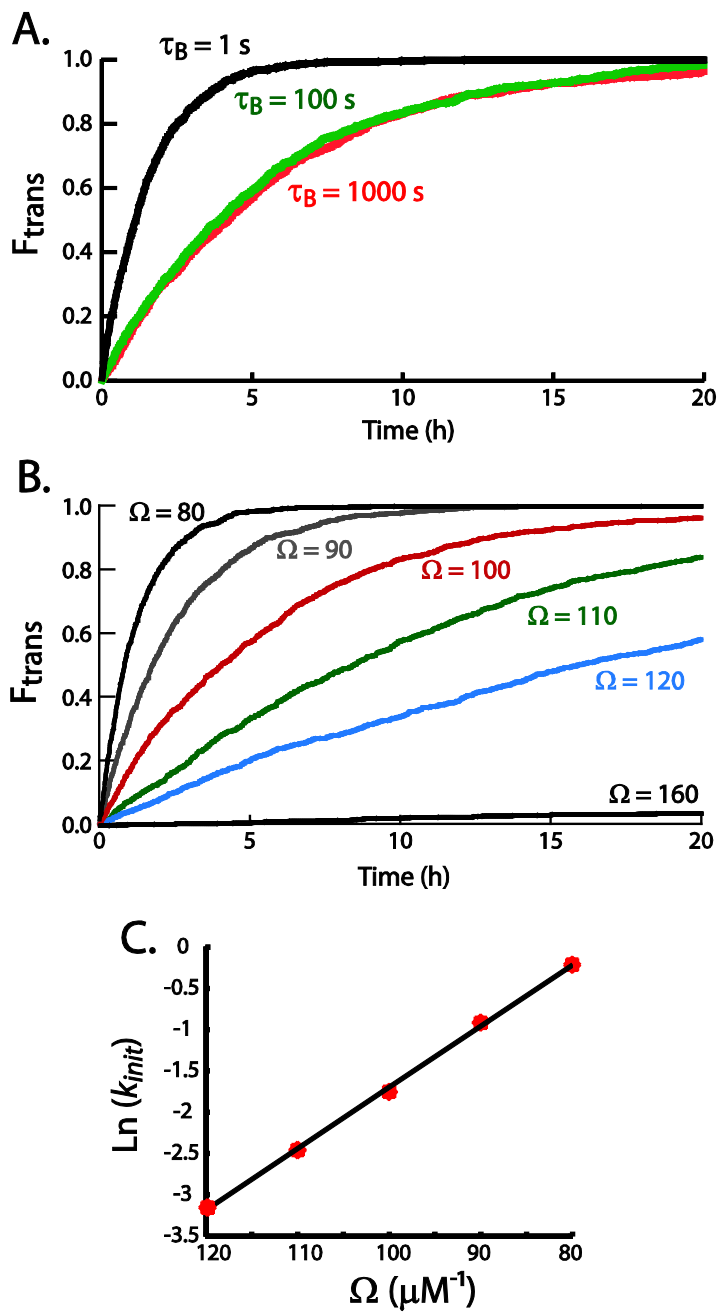

**Figure 6.** (Color Online) Robustness of bistability against fluctuations as a function of the time constant $\tau_B$ and of the volume factor $\Omega$. Other parameters are as in Fig. 5C. Simulations are initialized in the lower steady state of A and B. S is constant at 0.1. Each time course represents the evolution of the fraction of simulations that have transited to the upper state for an ensemble of 1,000 simulations. (A) Increase in stability with $\tau_B$. (B) Increase in stability with $\Omega$. Other parameters are as in (A), and $\tau_B$ is fixed at 100 s. (C) Linear least squares fit to points denoting the natural logarithm of the transition rate constant as a function of $\Omega$.

Using similar ensembles of simulations, Song et al. [41] demonstrated that for a bistable model of gene regulation, the logarithm of the mean first passage time (MFPT) from one steady state to another increased linearly with $\Omega$. Such a linear dependence for bistable kinetic models was predicted by Bialek [42]. For the autoactivation model, we tested this prediction by fitting exponential curves of the form $(1-\exp(-kt))$ to the time courses of Fig. 6B. In each of the ensembles of simulations characterized by different $\Omega$, the average initial rate at which simulations leave the lower state is proportional to the reciprocal of the MFPT. Thus, a plot of this rate vs. $\Omega$ is expected to be linear. This initial rate is the derivative of the exponential function evaluated at $t = 0$, which is simply k. Figure 6C verifies the linearity of the plot log (k) vs. $\Omega$, with k denoted by $k_{init}$. The points lie close to their linear least-squares fit. Intuitively, increasing $\Omega$ enhances the robustness of bistability because $\Omega$ corresponds to increasing average molecule numbers. With higher average molecule numbers, stochastic fluctuations in these numbers are relatively less significant, and have less effect on system dynamics.

**DISCUSSION**

Simple models of coupled feedback loops [2, 16] have proven fruitful for enhancing understanding of the advantages an organism can gain from a dual-time architecture. We have extended the analyses of Brandman et al. [2] and Zhang et al. [16] by 1) examining the dynamics conferred by multiplicative positive feedback in which the intermediate variables A and B multiply to increase the synthesis rate of the output variable $C_{out}$, and 2) examining whether the dual-time architecture of these models, with coupled fast and slow feedback loops, confers robustness of stable states against stochastic fluctuations in molecule numbers. With multiplicative feedback, the dual-time architecture conferred resistance to stimulus noise (Fig. 1B). Either a fast or slow turn on of $C_{out}$ production could be obtained, depending on parameters (Fig. 1B). A substantial range of bistability was obtained (Fig. 2A). With stochastic fluctuations in molecule numbers, robustness of bistable steady states was enhanced by the dual-time architecture. As the time constant of the slower positive feedback loop was increased, the average time required for molecule number fluctuations to destabilize a steady state increased rapidly (Fig. 2C).

The multiplicative variants of the models of Brandman et al. [2] and Zhang et al. [16] were studied because some of the specific dual-feedback systems noted by Brandman et al. [2] may be better described by multiplicative feedback. In muscle cell fate specification, the transcription factor MyoD is activated by the CDO Ig receptor and in turn upregulates CDO transcription [43]. This activation of MyoD is *via* enhanced dimerization. In a complementary feedback loop, MyoD activates transcription of Akt2 kinase, which in turn phosphorylates and further activates MyoD [44]. If muscle differentiation depends on having sufficient MyoD that is both dimerized and phosphorylated, then the feedback necessary for differentiation could be represented by Mod-B05-Mult with A = Akt2 kinase, B = CDO receptor, and $C_{out}$ = fully active MyoD. In B cell fate specification, the cytokine IL-7 appears to upregulate expression of the necessary transcription factor EBF, with EBF in turn upregulating transcription of the IL-7 receptor [45]. In a second feedback loop, EBF transcription is upregulated by another transcription factor, E2A [46]. If both feedback loops were required to produce sufficient active EBF to drive B cell differentiation, then this system might also be represented by Mod-B05-Mult with A = IL-7 receptor, B = E2A, and $C_{out}$ = EBF.

We also developed a similar model to represent the dual-time nature of coupled feedback loops postulated to be involved in LTP (Fig. 4). LTP induction and consolidation has been proposed to involve positive feedback in which MAPK indirectly enhances its own phosphorylation and activity [23]. Therefore, in the autoactivation model, the variables A and B could respectively represent the levels of active synaptic MAPK and total MAPK. LTP induction and consolidation has also been proposed to involve positive feedback in which CAMKII enhances its own phosphorylation and activity [26, 27]. Therefore, for modeling LTP, the variables A and B could respectively represent the levels of active synaptic CAMKII and total CAMKII. To apply the autoactivation model to LTP, a necessary assumption is that an increase in active synaptic kinase leads to an increase in total synaptic kinase. One way this could occur is if potentiated synapses, with a higher level of active kinase, recruited kinase molecules diffusing (or being transported) in neuronal processes. For CAMKII, there is evidence for such recruitment. NMDA receptor-dependent LTP induced by forskolin application is accompanied by a substantial increase in the amount of CAMKII in dendritic spines [47]. Application of NMDA also increases CAMKII in spines [48].

With the dual-time autoactivation model, rapid stimulus responses were obtained (Fig. 5) as was bistability (Fig. 4B). The dual-time architecture also stabilized steady states and responses against stimulus noise. Internal noise due to fluctuations in molecule numbers was simulated for the autoactivation model (Figs. 5-6). The dual-time architecture increased the robustness of bistability to internal noise. The time required for fluctuations to induce spontaneous escape from one of the two stable states to the other increased substantially as the B feedback loop was made slower. However, the increase in stability saturated when $\tau_B$ was ~ 100 times longer than the fast loop time constant $\tau_A$ (Fig. 6A). The volume factor $\Omega$ exerted a much stronger effect on the stability of steady states (Fig. 6B).

For the autoactivation model, the state of elevated kinase activity develops resistance to reversal by brief stimuli, over the course of a few hours (Figs. 5A and 5C). This resistance is due to elimination of the stable lower state by a slow increase in total kinase amount (Fig. 5B). These dynamics suggest an explanation for the empirical development of resistance of LTP to reversal. Initially, synaptic potentiation may rely on an increase in active kinase which could be rapidly reversed by dephosphorylation of kinase (MAPK or CAMKII). Subsequently, a slow increase in total synaptic kinase may result in persistently elevated kinase activity and synaptic strength that is resistant to reversal. Empirically, following induction of LTP *in vitro*, subsequent electrical stimuli can reverse LTP only when applied within ~ 1 h [35, 36]. Similar dynamics are observed *in vivo* [49, 50]. The time course of ~1 h *in vitro* and *in vivo* could correspond in the model to the time required for increases in the total amounts of synaptic proteins. Empirically, the late phase of LTP (L-LTP) does require both translation and transcription [51, 52].

A much longer stimulus might still reverse established L-LTP by decreasing the level of total kinases and other synaptic proteins. In the model, prolonged inhibition of protein synthesis would sufficiently decrease B to re-establish stability of the lower state of active kinase A. Indeed, a sufficiently large decrease in B is seen to eliminate the stable upper state of A for S = 0, after which A spontaneously falls to the lower state. For the parameters of Fig. 5B and for S = 0, the upper state of A is lost below B = 0.5. Empirically, prolonged inhibition of glutamatergic neurotransmission by inducible NMDA receptor knockout eliminates established LTM [53, 54]. This LTM elimination plausibly corresponds to reversal of established LTP due to prolonged block of activity-dependent protein synthesis.

The slow feedback loop in the autoactivation model postulates that an increase in active kinase (variable A) leads to an increase in total kinase (variable B). Activation of transcription, subsequent to kinase activation, may be one mechanism that enhances levels of total synaptic kinase. In mammalian cells, MAPK can phosphorylate and activate ribosomal S6 kinase (RSK) [55] and RSK can phosphorylate and activate the transcription factor cAMP response element binding protein (CREB) [56]. Phosphorylation and activation of CREB in neurons correlates with recruitment of those neurons into a long-term memory trace [57], plausibly by strengthening synapses to or from these neurons. It is plausible that additional positive feedback in which CREB enhances its own transcription plays a role in consolidation of late phases of LTP and LTM. Mammalian *creb* has cAMP response element (CRE) enhancer sequences in its promoter [58]. CREB activates transcription *via* binding to CREs. Thus, activation of CREB might initiate positive feedback based on *creb* autoregulation.

The autoactivation model (Fig. 4A) may also describe aspects of LTF in invertebrates. In *Aplysia*, MAPK and PKA are activated during LTF [59, 60]. Inhibition of MAPK blocks LTF [61]. As suggested for LTP, positive feedback involving persistent MAPK phosphorylation might contribute to LTF. In *Aplysia*, activation of the CREB1 transcription factor is necessary for LTF [62]. *Aplysia creb1* has a CRE [63], and is activated by CREB1 [15]. The positive feedback loop in which CREB1 enhances transcription of *creb1* was recently shown to be important for consolidation of LTF [15].

Additional positive feedback loops may contribute to LTF. In one putative loop, enhanced transcription of the ApTBL gene product increases levels of TGF-β growth factor, which acts through receptors to further activate MAPK, phosphorylate transcription factors, and maintain enhanced transcription [14]. In a second proposed loop, protein kinase A (PKA) acts to induce expression of *Aplysia* ubiquitin hydrolase (Ap-uch) [64, 65]. Ap-uch regenerates free ubiquitin, prolonging PKA activity by promoting proteosome-dependent degradation of the regulatory subunit of PKA [66]. Proteosome-dependent protein degradation is also necessary for mammalian LTP [67]. Activation of translation might also enhance levels of kinases such as MAPK. Aggregation of a translational activator, cytoplasmic polyadenylation element binding protein (CPEB), was proposed to maintain enhanced translation at synapses that have undergone LTF, thereby maintaining LTF [68].

The above data suggest that, in both mammals and invertebrates, feedback loops involving regulation of transcription by CREB and enhanced proteosome-dependent protein degradation may play important roles in the formation of LTM. The similarity of biochemical pathways involved in LTM in evolutionarily divergent animals suggests that generic models similar to those studied here, with simple representations of fast and slow feedback loops, may help in understanding memory formation in a broad range of organisms.

**ACKNOWLEDGEMENTS**

We thank M. Byrne for assistance with the preparation of the manuscript and the illustrations and Y. Zhang and L. Zhou for comments on an earlier draft of the manuscript. This work was supported by National Institutes of Health Grant P01 NS038310.